\documentclass{jpp}
\usepackage{graphicx}
\usepackage{epstopdf, epsfig}
\usepackage{hyperref}
\usepackage{amsmath}
\usepackage{color}
\usepackage{subfigure}

\usepackage{bm}

\shorttitle{The effect of shaping on RFP dynamics}
\shortauthor{R. Chahine, K. Schneider and W.J.T. Bos}

\title{The effect of shaping on turbulent dynamics in RFP simulations}

\author{Robert Chahine\aff{1}
 \corresp{Current address: Rio Tinto Aluminium, Voreppe, France.},
  Kai Schneider\aff{2}
 \and Wouter J.T. Bos\aff{1}
}

\affiliation{\aff{1}Univ Lyon, CNRS, Ecole Centrale de Lyon, INSA Lyon, Univ Claude Bernard Lyon 1, LMFA,
UMR5509, 69340 Ecully, France
\aff{2}I2M, Aix-Marseille Universit\'e, CNRS, Centrale Marseille, 13453 Marseille, France}

\begin{document}

\maketitle

\begin{abstract}
We study the influence of the shape of the plasma container on the dynamics of the Reversed Field Pinch (RFP). The geometries we consider are periodic cylinders with elliptical and circular-shaped cross-sections. Numerical simulations of fully nonlinear visco-resistive magnetohydrodynamics are carried out to illustrate how the plasma dynamics are affected by shaping. {It is shown that independent of the plasma shape, the quantity $\beta$, comparing the hydrodynamic pressure to the magnetic pressure, decreases for increasing values of the Lundquist number, but the pressure gradient fluctuations remain roughly constant, when compared to the Lorentz force. Different elliptical shapes of the cross-section of the domain lead to the excitation of different toroidal (or axial) magnetic and dynamic modes.
Furthermore, it is found that in a geometry with circular cross-section a significant local poloidal angular momentum is observed, absent in the geometries with elliptical cross-section.} 
Since the confinement is dominantly determined by plasma movement, and the dynamics of the velocity and magnetic field are modified by the modification of the geometry, shaping can thus affect the performance of RFP-devices.
\end{abstract}

\section{Introduction}

Tokamaks and Reversed Field Pinches (RFPs) are toroidal fusion plasma devices with a similar magnetic geometry. In both types of reactors the combination of an imposed toroidal magnetic field combined with a poloidal magnetic field, associated with an induced toroidal current, result in helical magnetic field lines, around the toroidal axis. The difference between tokamaks and RFPs is the strength of the toroidal magnetic field, which needs to be much larger than the poloidal field in tokamaks, whereas it is of the same order of magnitude in RFPs. This requirement is due to threatening magnetohydrodynamic (MHD) instabilities, or disruptions, in tokamaks \citep{biskamp}, which lead to loss of confinement and possibly damage the reactor.

RFPs work in this unstable regime, but take advantage of the nonlinear saturation of the instability, thereby bypassing the risk of disruptions and avoiding the need of a very strong (and costly) toroidal magnetic field. Whereas in early research on the RFP the unstable character was seen as a drawback for fusion, it has become increasingly clear that the self-organization of the RFP is actually an asset to reach self-sustained fusion. Indeed in the 2000s, quasi-single-helicity (QSH) states were detected within turbulent flows in the RFX experiment \citep{escande2000,martin_2000,martin_2003}. These states are characterized by the appearance of a quiescent helical structure in the plasma core, which improves the plasma confinement \citep{frassinetti_2006,terranova_2007,wyman_2008}. Later studies showed that the persistence of these QSH states and the appearance of a Single-Helical Axis at high current regimes \citep{piovesan_2009} can be increased by applying helical magnetic perturbations \citep{piovesan_2011,piovesan_2013}. 

These results motivated part of the fusion community to reconsider the RFP as a serious candidate for nuclear fusion \citep{lorenzini_2009}. {Even though the road to ignition, i.e., self-sustained fusion is still long for RFPs, and they might not be the most promising geometry to attain the eventual goal of fusion research, they constitute an interesting example of plasma self-organization and their investigation can help to develop ideas which might be useful for other reactor-geometries.}
Recently the RFP has therefore received renewed attention, focusing on various aspects, such as fast ion transport \citep{bonfiglo2019fast}, ion-temperature-gradient modes \citep{li2019effects}, sophisticated three-dimensional equilibria \citep{qu2020stepped} and reconnection \citep{momo2020phenomenology}. A recent review, centered on the Madison Symmetric Torus experiment, can be found in \citet{sarff2020reversed}.


 
{Acting on the RFP magnetic field to improve its confinement properties is evidently an important path to investigate. For instance,} applying helical magnetic perturbations seem a promising way to affect the self-organized state in an RFP \citep{bonfiglio2013experimental}.   {Such perturbations were shown to be able to influence the shape of the plasma by imposing their helical pitch to the plasma \citep{veranda2017magnetohydrodynamics}}. Another obvious way  to affect the dynamics would be to change directly the global shape of the plasma. The optimization of the confinement quality for toroidal fusion plasmas by changing the plasma shape has been the subject of many studies,
 in particular for tokamaks. For instance, it has been shown
 that shaping has a beneficial effect on the $\beta$ limits of tokamaks \citep{troyon1984}, and increases the total plasma current $I$ in the case of elliptic cross-sections, yielding thus a better confinement. Furthermore, shaping of a tokamak cross-section can lead to qualitative differences in the  plasma flow-patterns \citep{morales2012intrinsic,oueslati2020breaking}. Investigations on the influence of shaping on the confinement properties of RFPs are, however, relatively scarce. The literature on RFPs contains some rare examples of experimental observations\citep{almagri_1987,oomens_1990}, and numerical investigations where 
 two-dimensional equilibrium studies were carried out in order to investigate the shaping effect on RFP plasmas \citep{paccagnella1991,guo2013}. Their work led
 to the conclusion that shaping does not bring an advantage to the plasma dynamics in RFPs and is even destabilizing in the case in which the poloidal cross-section is elongated.
 These studies focused on the {linear} stability properties of RFPs, but did not consider the fully developed nonlinear dynamics.

 Here we proceed one step further in the investigation of the effect of changing the shape of the cross-section of RFPs by considering the 
 fully nonlinear dynamics within a resistive fluid description. More precisely, we investigate the effect of elongation of the poloidal
 cross-section on plasmas in driven incompressible MHD in cylindrical geometry. We thereto perform direct numerical simulations using
 a three-dimensional pseudo-spectral solver \citep{morales_jcp}. We consider the simplified case where the torus is modeled by a straight periodic cylinder. 
 The choice of this simplification is justified as  follows. In \citet{morales_ppcf}, we compared the straight-cylinder approach to fully toroidal simulations. We showed that most of the qualitative features remained unchanged. The most significant change was the appearance of a toroidally invariant mode, the influence of which we do therefore necessarily neglect in the present work. It is true that considering the effect of curvature on the dynamics of RFPs could be interesting, but we aim at the understanding of the two effects (curvature and shape of the cross-section) independently in order to pinpoint the most important physical effects, before considering their possible interplay. Furthermore, in \citet{paccagnella1991} the influence of curvature was considered with respect to the stability properties of RFPs and its effect was shown to be minor. 
 
In the present work it is shown that elongation of the cross-section has a significant effect on the pressure-statistics in  the plasma and on the turbulence properties in general. In particular, assessing the value of the $\beta$-parameter, and its recent generalization to take into account the turbulent pressure-gradients, illustrates how the dynamics are influenced by the shaping of the geometry. {We also reveal the presence of non-negligible local poloidal angular momentum, when the cross-section is circular.}


 
The remainder of this article is organized as follows. Sec.~\ref{sec:set-up} presents the governing equations, recalls briefly the numerics and the relevant physical parameters. Numerical results on the influence of shaping on the reversal-properties,  pressure-statistics, spectral characteristics and local angular momentum are investigated in Sec.~\ref{sec:results}. Sec.~\ref{sec:conc} concludes the investigation.

\section{Equations, numerical methods and parameters}\label{sec:set-up}

\subsection{Visco-resistive MHD equations and geometry}

 In the present work, we consider a plasma characterized by constant, and uniform permeability $\mu$, permittivity $\epsilon$ and conductivity $\sigma$.  The more complicated case of non-uniform conductivity, was considered in \cite{futatani_pop}.  In the magnetohydrodynamic (MHD) description that we consider, the governing equations are the incompressible Navier-Stokes equations including the Lorentz force, and the induction equation.
 Normalizing these equations by the Alfv\'en velocity $C_A=B_0/\sqrt{\rho\mu}$, a reference magnetic field $B_0$ and a conveniently chosen 
 lengthscale $\mathcal{L}$ leads to the following expressions for the evolution of the velocity $\bm u$ and magnetic field $\bm B$, 
    \begin{equation}\label{mhd_1}
  \frac{\partial \bm u}{\partial t} + \bm u\cdot\nabla\bm u = -\nabla P  + \bm j\times\bm B + \frac{P_m}{S}\nabla^2\bm u ,
 \end{equation}
  and
 \begin{equation}\label{mhd_2}
  \frac{\partial \bm B}{\partial t} = \nabla\times(\bm u\times\bm B) + \frac{1}{S}\nabla^2\bm B, 
 \end{equation}
 where $P_m$ is the magnetic Prandtl number $P_m=\nu \mu\sigma$, $\nu$ the viscosity, $S$ the Lundquist number (defined below)  and $\rho=1$ the density.
 The current density is given by
  \begin{equation}
  \bm j=\bm\nabla\times\bm B .
 \end{equation}
 The velocity field $\bm u$ and the magnetic field $\bm B$ are both divergence free,
\begin{eqnarray}
\label{eq:divu}
 \bm\nabla\cdot\bm u = 0,\\
 \bm\nabla\cdot\bm B= 0.
\end{eqnarray}
The incompressibility condition (\ref{eq:divu}) allows obtaining the pressure $P$ from the velocity field by taking the divergence of equation (\ref{mhd_1}) and solving the resulting Poisson-equation.  {The pressure plays thereby in this system the role of a Lagrange multiplier, enforcing incompressibility of the velocity field}. We think it is important to retain this feature (incompressibility) in the dynamics unlike in a number of previous investigations of RFPs (e.g. \citet{Cappello_1996,cappello_2000,richardson2010control,veranda2017magnetohydrodynamics,futch2018role}), where the pressure was entirely neglected invoking low-$\beta$ dynamics. Indeed, in some recent RFP-investigations the pressure is retained \citep{mizuguchi2012modeling}, and recently we focused on the importance of the pressure dynamics \citep{chahine2018role}. In particular we illustrated in MHD simulations in cylindrical geometry that, in order to understand the dynamics of the plasma, the important quantity to monitor is not the pressure, but its gradient. We therefore introduced a quantity $\beta_\nabla$, which compares the influence of the pressure gradient to the magnetic effects ($\bm j\times\bm B$) acting on the momentum balance. {The definition of this quantity and its interpretation will be given in section \ref{sec:beta}.}

In the present work we thus take into account the influence of pressure on the dynamics, but we neglect all compressibility effects. Note that imposing incompressibility was shown to diminish the reversal of the magnetic field \citep{finn1992single}, and this will thus necessarily be the case in the present investigation. We note here that the resistivity profile can also influence the reversal \citep{bonfiglio2016impact}. The combined influence of shaping, compressibility and non-uniform resistivity would constitute an interesting perspective, but would complicate disentangling the different effects.

\begin{figure}
\begin{center}
\hspace*{10pt}
\includegraphics[width=0.3\columnwidth]{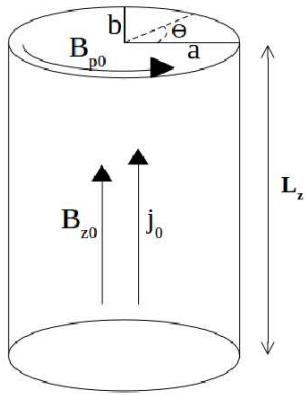}
\caption{Sketch of the cylindrical geometry and imposed magnetic field and current density. \label{fig:sketch}}
\end{center}
\end{figure}

In Fig.~\ref{fig:sketch} we illustrate the considered geometry and we indicate the direction of the imposed current density and magnetic field. Initially, in the plasma a uniform current density $j_0$ in the $z$-direction and an axial
magnetic field $B_{z0}$ are imposed, resulting in a helically shaped magnetic field. The current density $j_0$ will
induce an elliptical {poloidal} magnetic field $B_{p0}$ parallel to the elliptic boundaries.  At later times the magnetic field will reorganize through an interplay with the velocity field, and the total magnetic field will then consist of $B_{z0}$ and $B_{p0}$ plus the self-induced contributions.
At the boundaries 
the velocity is imposed to be zero and the magnetic field is parallel to the boundaries. The value of the poloidal parallel magnetic
field at the boundary is fixed and its value is  determined by $j_0$. The expression of $B_{p0}$ in cylindrical coordinates reads,
\begin{equation}\label{br_ell}
 B_r=-\frac{1}{2}j_0r c\sin(2\theta) 
\end{equation}
\begin{equation}\label{bt_ell}
 B_\theta=\frac{1}{2}j_0 r(1-c\cos(2\theta))
\end{equation}
with $c$ the ellipticity which can be expressed as a function of the ellipse's major semi-axis $a$ and minor semi-axis $b$, i.e., 
\begin{equation}
 c=\frac{a^2-b^2}{a^2+b^2}.
\end{equation}
 Note that the coordinates we use are cylindrical and not elliptical coordinates so that only in the case of the circle the radial vector $\bm e_r$ is everywhere perpendicular to the boundary.


%

\subsection{Numerical methods}

Equations \eqref{mhd_1} and \eqref{mhd_2} are solved using a 
pseudo-spectral method in a periodic domain of size $\pi\times\pi\times 8\pi$ with $64\times64\times512$ grid points. The aspect ratio of the physical domain containing the plasma is $L_z/2\pi  b =4$. Spatial derivatives are evaluated 
in Fourier space and multiplications are computed in physical space. To avoid aliasing errors, i.e., the production of small scales
due to nonlinear terms which are not resolved on the grid, the velocity and magnetic fields are dealiased at each time step by truncating 
its Fourier coefficients using the 2/3 rule \citep{aliasing}. Using the incompressibility condition of the fluid, the pressure term 
can be eliminated by solving a Poisson equation. A semi-implicit time-advancing scheme of Adams-Bashforth type is used to solve the equations, with exact integration of the dissipative and magnetic 
diffusion terms. Boundary conditions are imposed using a volume penalization method in order to build the cylindrical domain. Detailed description and validation
of the method can be found in \citet{morales_jcp}, and an application of the method to investigate RFPs in toroidal domains is 
reported in previous work \citep{morales_ppcf,futatani_pop}. 
We have verified by assessing the high-wavenumber range of the kinetic and magnetic energy spectra that the resolution for all simulations was sufficient to resolve all down to the smallest dynamical flow scales.


Simulations are started from random initial conditions and the dynamics are observed to evolve towards an initial-condition independent statistically stationary state. To obtain the results presented in Sec.~\ref{sec:results}, the equations are integrated for $10^4 \tau_A$ Alfv\'en times, with 
$\tau_A = \mathcal{L}/C_A$. 
The results present presented in the following are
all obtained during the statistically stationary state.


\subsection{Shaping parameters}

In the present investigation we focus on the influence of the shape of the cross-section on the confinement properties of the
plasma. The parameters should be carefully chosen to disentangle the effect of changing the geometry from the effect of 
changing other control parameters. Considering a periodic cylinder instead of a torus is motivated by this 
attempt to reduce the number of control parameters to a strict minimum. Even in this simplified geometry,
the way in which the parameters are varied is not unique. For instance, if the same 
toroidal current-density $J_z$  is chosen for two geometries, the mean current $I_z$ will be the same, only if the 
surface $A$ of the cross-section is kept constant, a condition which we will impose. This will also lead to equal values of the toroidal 
magnetic flux $\psi=B_z A$, for a given imposed toroidal magnetic field $B_z$.
\\The poloidal magnetic $B_p$ field is computed from the current density. Its reference value is evaluated as an average over 
the circular, or elliptic boundary. Necessarily, keeping the surface $A$, $J_z$ and $B_z$ fixed, the average value 
$\overline B_p$ varies when changing the shape of the cross-section (the bar indicates a boundary average). 
Therefore, the pinch-ratio, defined as
\begin{equation}
\Theta = \frac{\overline{B_p}}{\langle B_z\rangle},
\end{equation}
where the brackets denote a volume average, depends on the value of the ellipticity $c$. 

{An important parameter in non-ideal MHD is the Lundquist number, which is defined as
\begin{equation}
S=\frac{2 C_A b}{\lambda},
\end{equation}
where the Alfvén velocity is based on the poloidal field strength, $C_A = B_p/\sqrt{\rho\mu}$ and with the magnetic diffusivity $\lambda=(\mu\sigma)^{-1}$ and the (minor) diameter $2b$ as reference quantities. }
We have chosen $b$, rather than $a$, since the smallest minor radius which will probably determine the confinement quality. 
Imposing the same value of $S$ for 
different values of the ellipticity allows determining the value of $\lambda$. In all our simulations the value of the 
magnetic Prandtl number is chosen unity.

\section{Results}\label{sec:results}
\subsection{F-$\Theta$ dependence}
The imposed magnetic field in RFPs is unstable for large values of $\Theta$ and $S$, and it will form a dynamic helical structure 
with a certain amount of chaotic or turbulent motion superimposed. 
\\The modification of the magnetic field can be quantified by the field reversal parameter $F$, representing the normalized toroidal field at the boundary,
\begin{equation}
F = \frac{\overline{B_z}}{\langle B_z\rangle}.
\end{equation}
As the current increases, the kink instability increases, leading to the decrease of the toroidal magnetic field at the boundary, so that $F$ decreases as a function of $\Theta$. This behavior can be qualitatively predicted by Taylor's theory \citep{taylor_1974} and more sophisticated theories allow to improve this agreement \citep{reiman_1980,reiman_1981,taylor_1986}.
In studies \citet{paccagnella1991,guo2013} based on two-dimensional equilibrium equations, it was shown that shaping 
does not alter the $F$-$\Theta$ curve. 


\begin{figure}
\begin{center}
\includegraphics[width=.6\textwidth]{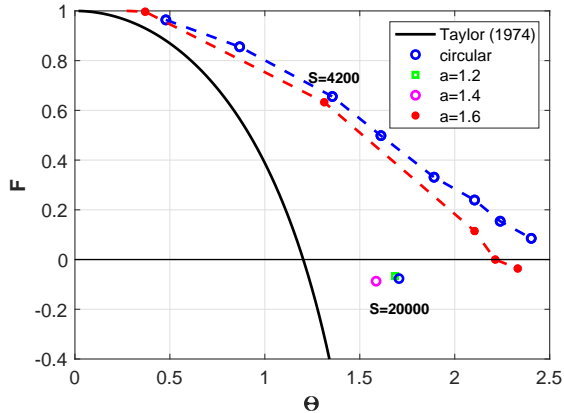}
\caption{Field reversal parameter $F$ as a function of the pinch parameter $\Theta$ for cylinders with circular and elliptic cross-section. Also shown is Taylor's prediction \citep{taylor_1974} for reference.}
 \label{ftheta}
 \end{center} 
\end{figure}

In order to be able to systematically assess the  $F$-$\Theta$ dependence, we carry out preliminary simulations for cylinders of small aspect ratio 
${L_z}/{2\pi b}\sim 2$, an ellipticity $a=1.6$ and moderate Lundquist number $S\sim 4200$. We compare with Taylor's prediction \citep{taylor_1974} and simulations with circular cross-section. Figure \ref{ftheta} shows the results of the field reversal parameter $F$ versus $\Theta$, which are in reasonable agreement with the two previous studies \citep{paccagnella1991,guo2013}.  In these references it was shown that shaping had a small destabilizing effect for large curvature, but the $F-\Theta$ curve was unaffected.  Indeed, the two geometries 
yield roughly the same behavior.

These moderate-Lundquist simulations allow a parametric investigation of the $F$-$\Theta$ dependence for two different ellipticities. Carrying this out for higher values of $S$ would  require substantially more computational resources. In the remainder of this investigation we will consider, at a higher value of the Lundquist number, three different ellipticities, but focusing on only one $F$-$\Theta$ value for each geometry. 
We have thereto performed simulations for values of the Lundquist number ranging up to $S \approx 2\cdot 10^4$ and a larger aspect ratio $L_z/2\pi b = 4$. {These higher values of the Lundquist number are still several orders of magnitude smaller than those in experimental RFPs, which are currently out of reach using precise numerical schemes.} %
We consider three shapes, {\it i.e.} cylindrical devices with different cross sections: a circle with radius $a=1$, an ellipse with  $a=1.2$ $b=0.83$ and an ellipse with $a=1.4$ and $b=0.714$. 
In the statistically steady states considered at this higher Lundquist number, both the $F$ and $\Theta$ parameter fluctuate around their steady value, which is also plotted in Fig.~\ref{ftheta}. {It is observed that these results are in the $F-\Theta$ plane situated closer to Taylor's prediction, which is obtained using statistical mechanics of ideal MHD \citep{taylor_1974}. We do not know whether higher values of $S$ will allow to approach this theoretical prediction even closer. }
The fluctuations are for $\Theta$ inferior to $0.5\%$. The absolute value of the fluctuations of the $F$-parameter are of order $0.01$ (for instance fluctuating between $F=-0.07$ and $F=-0.082$ for the circular geometry).

We observe thus that the effect of shaping is small on the global behavior of the magnetic field, as characterized by $F$ and $\Theta$. The present results indicate that the influence of the Lundquist number is significantly larger than that of shaping in the considered range of parameters.
The reversal parameter is a global parameter and does not give insight into the fine 
structure of the dynamics. It is this fine structure, constituted by the nonlinear interplay of a large 
number of modes which will determine the confinement quality of a reactor.
The modification of the fine structure is in the following assessed by evaluating pressure gradients and the modal behavior of the flow.

\subsection{The role of the pressure: $\beta$ and gradient-$\beta$. \label{sec:beta}}



\begin{figure*}
\hspace{-1cm}
\includegraphics[width=0.45\columnwidth]{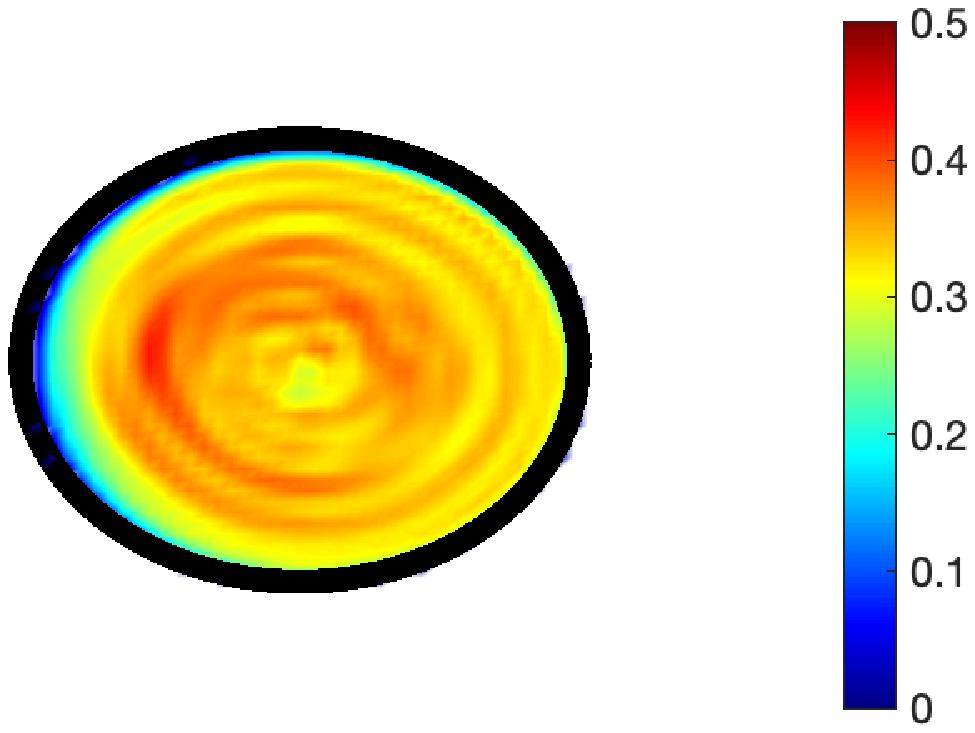}~\hspace{-2.2cm}
\includegraphics[width=0.45\columnwidth]{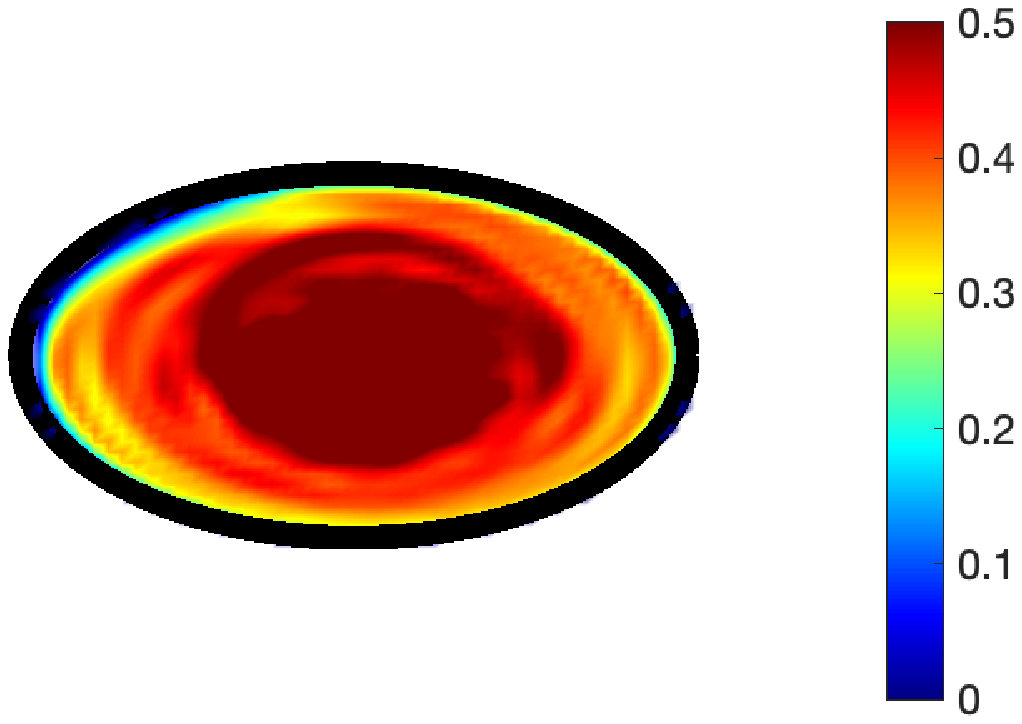}~\hspace{-2cm}
\includegraphics[width=0.45\columnwidth]{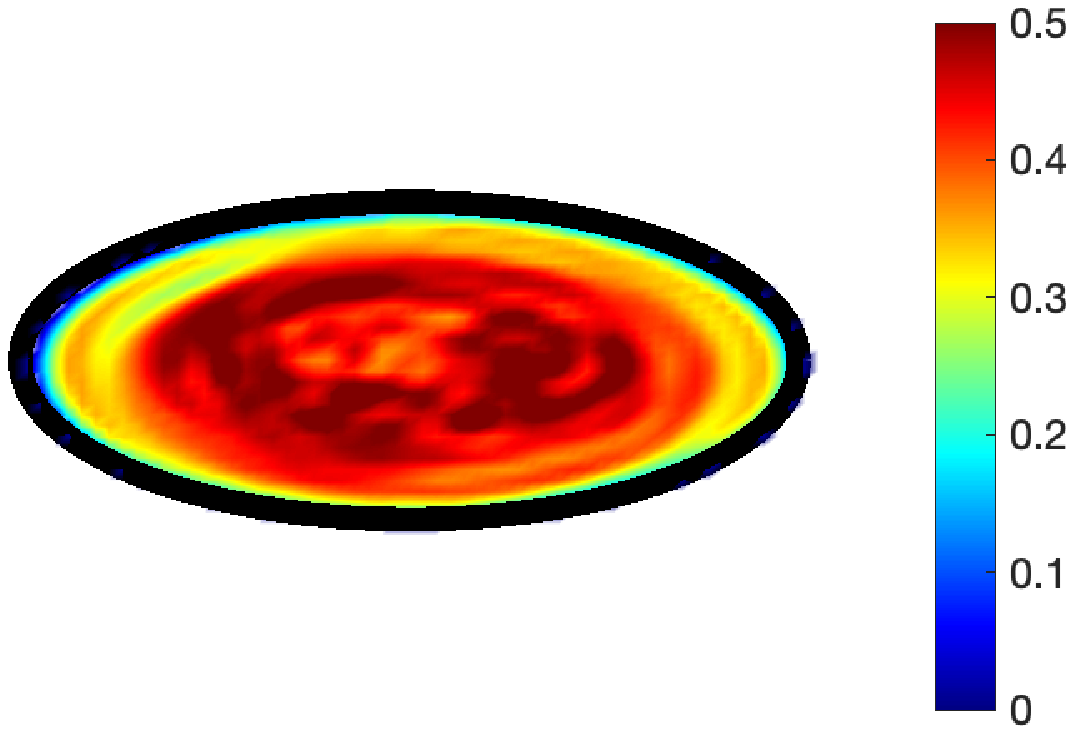}~~~~~~
\caption{Instantaneous visualizations of the pressure field for $S \approx 2 \cdot 10^4$ in cross-sections through the cylinders for the three different ellipticities. From left to right, $a=1$, $a=1.2$, $a=1.4$. 
 \label{fig:visu_beta} }
\end{figure*}

\begin{figure*}
\centering
\includegraphics[width=0.6\columnwidth]{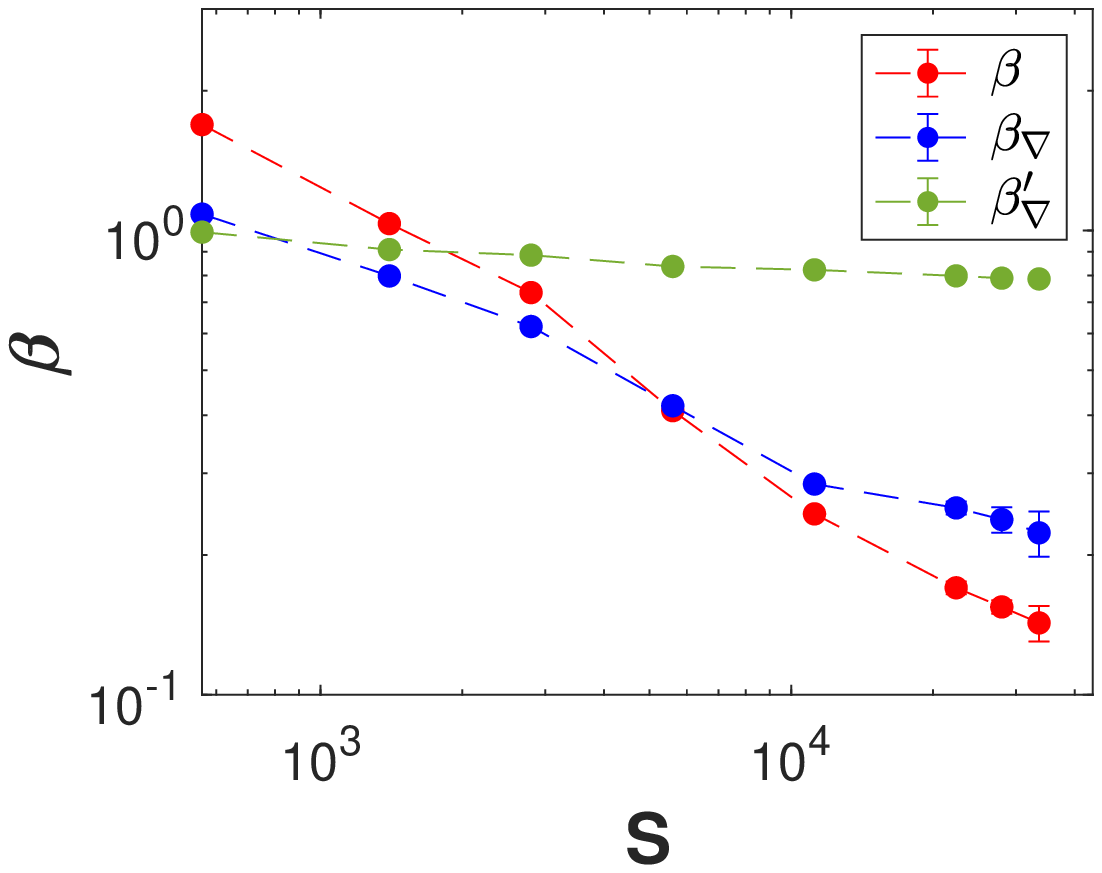}\\
\includegraphics[width=0.6\columnwidth]{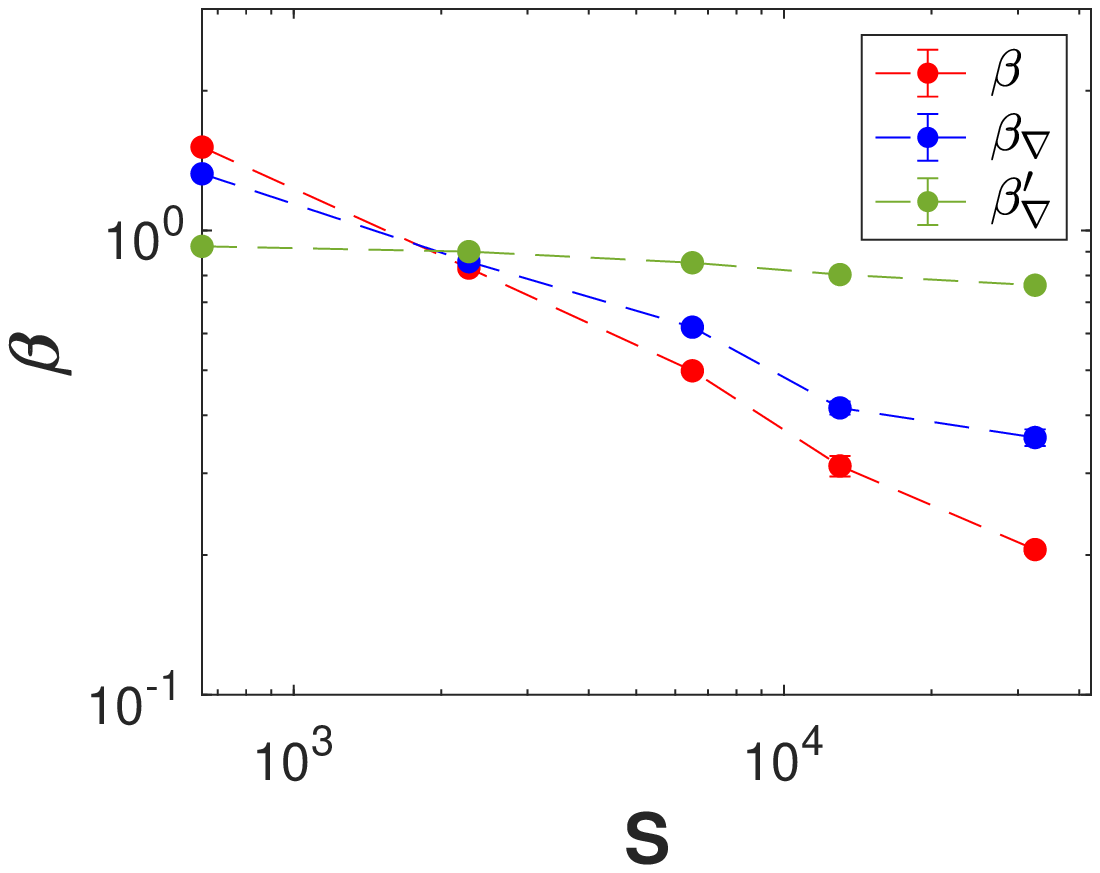}\\
\includegraphics[width=0.6\columnwidth]{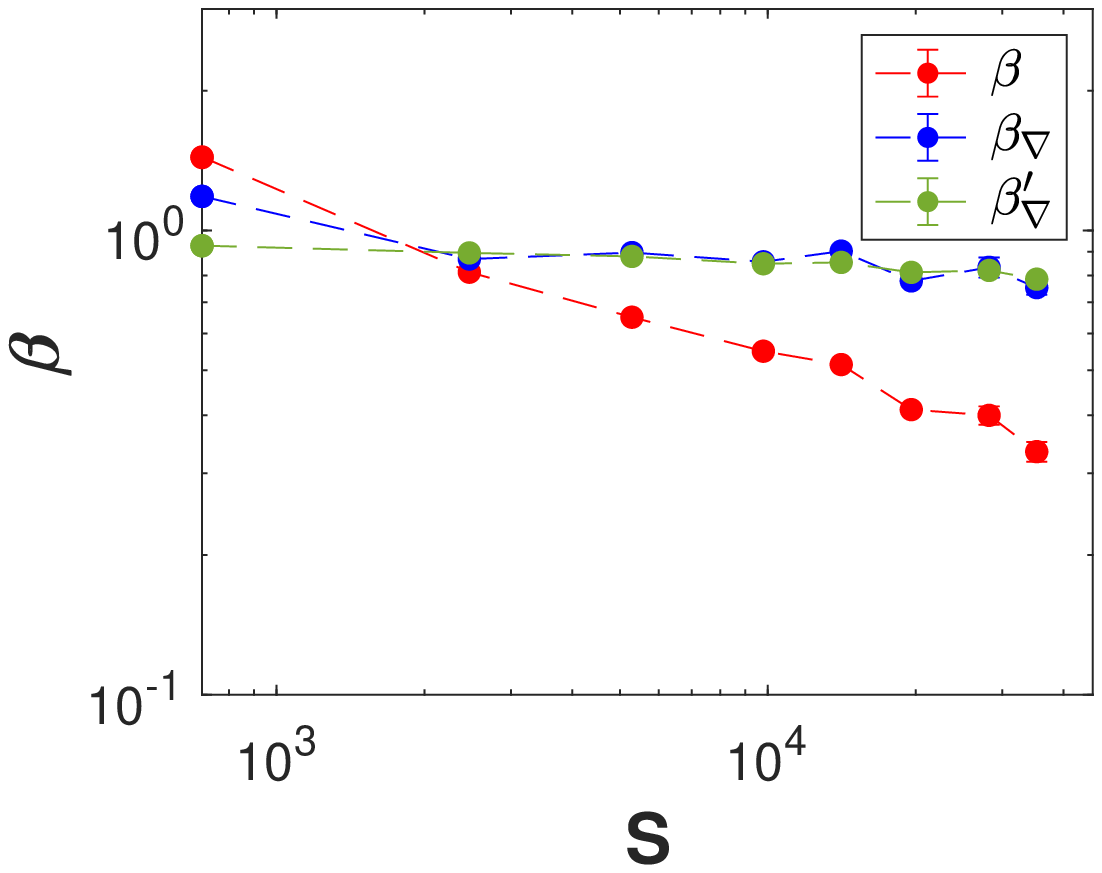}
\caption{Lundquist-number dependence of three alternative definitions of $\beta$ in cylinders with ellipticity (a) $a=1$, (b) $a=1.2$, (c) $a=1.4$. Whereas the pressure based $\beta$ is strongly decreasing with $S$, the gradient based value $\beta'_\nabla$ is roughly constant. \label{fig:beta}}
\end{figure*}

{Clearly, in fusion-research the plasma pressure plays a major role. In the present investigation, we do not focus on the confinement quality of the system, but rather on the dynamics and the velocity field.} Therefore we will assess the influence of shaping on the pressure and its influence on the dynamics. In Fig.~\ref{fig:visu_beta}, we show visualizations of the pressure fluctuations in cross-sections of the domains at the highest considered Lundquist number. Spatial fluctuations of the pressure are observed in all geometries.  In particular in the elliptic cases the strongest values of the absolute pressure are situated in the center of the domain and lower pressure values are observed close to the wall. In the circular case the value of the pressure is smaller than for the elliptic cases. 
We will now focus on different measures of the influence of pressure on the dynamics.

\begin{figure}
\hspace{-1.cm}
\includegraphics[width=.4\textwidth]{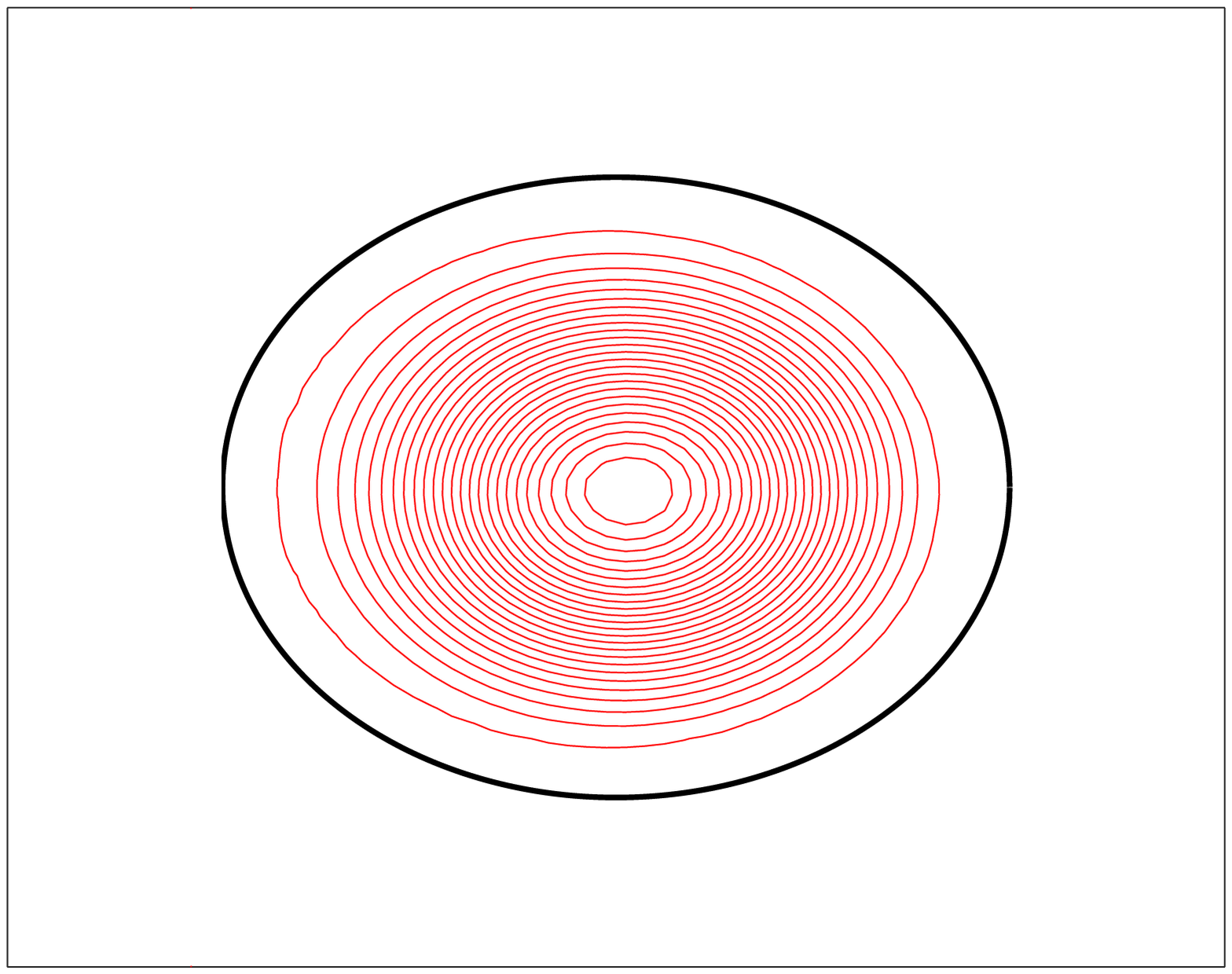}~\hspace{-0.6cm}
\includegraphics[width=.4\textwidth]{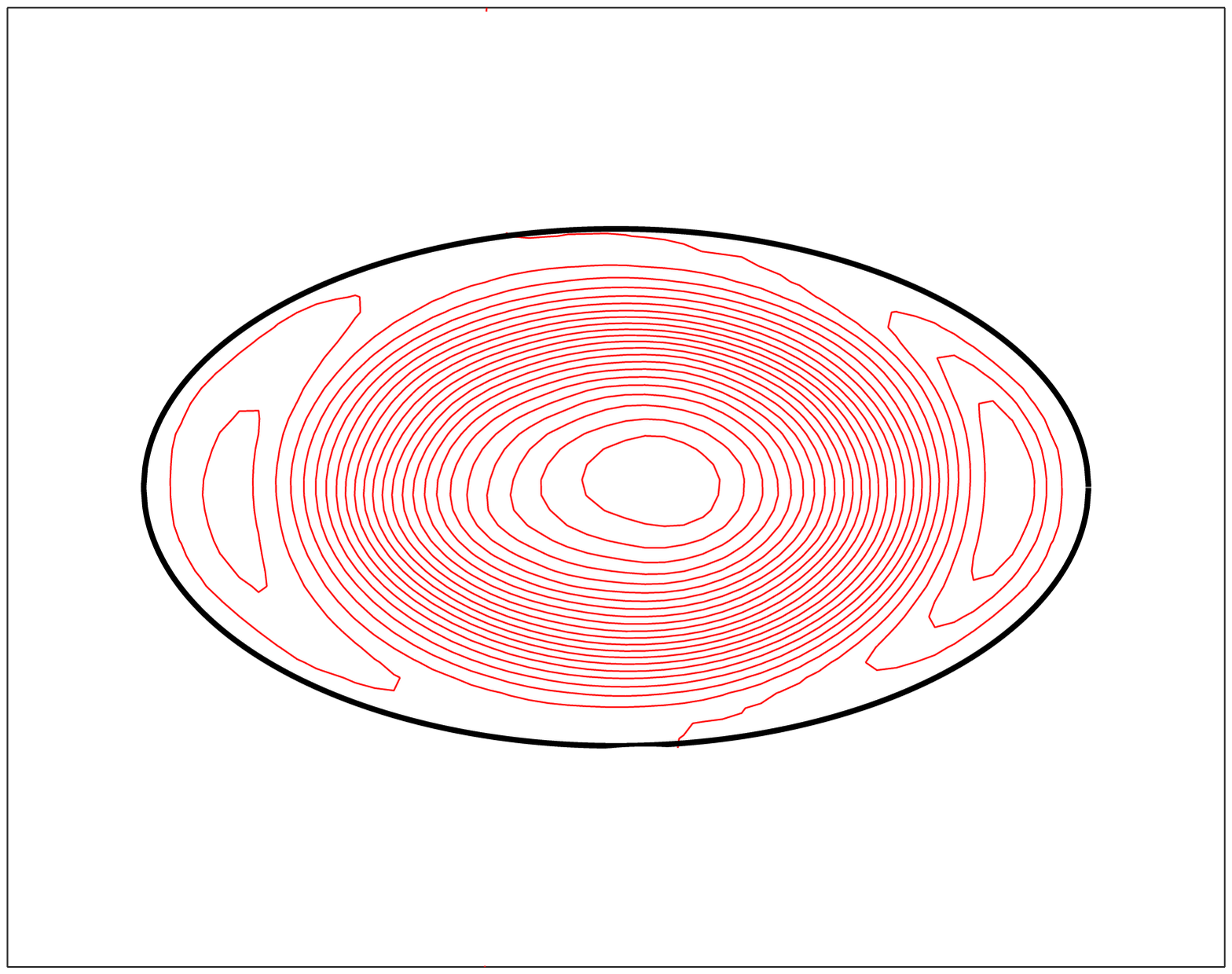}~\hspace{-0.6cm}
\includegraphics[width=.4\textwidth]{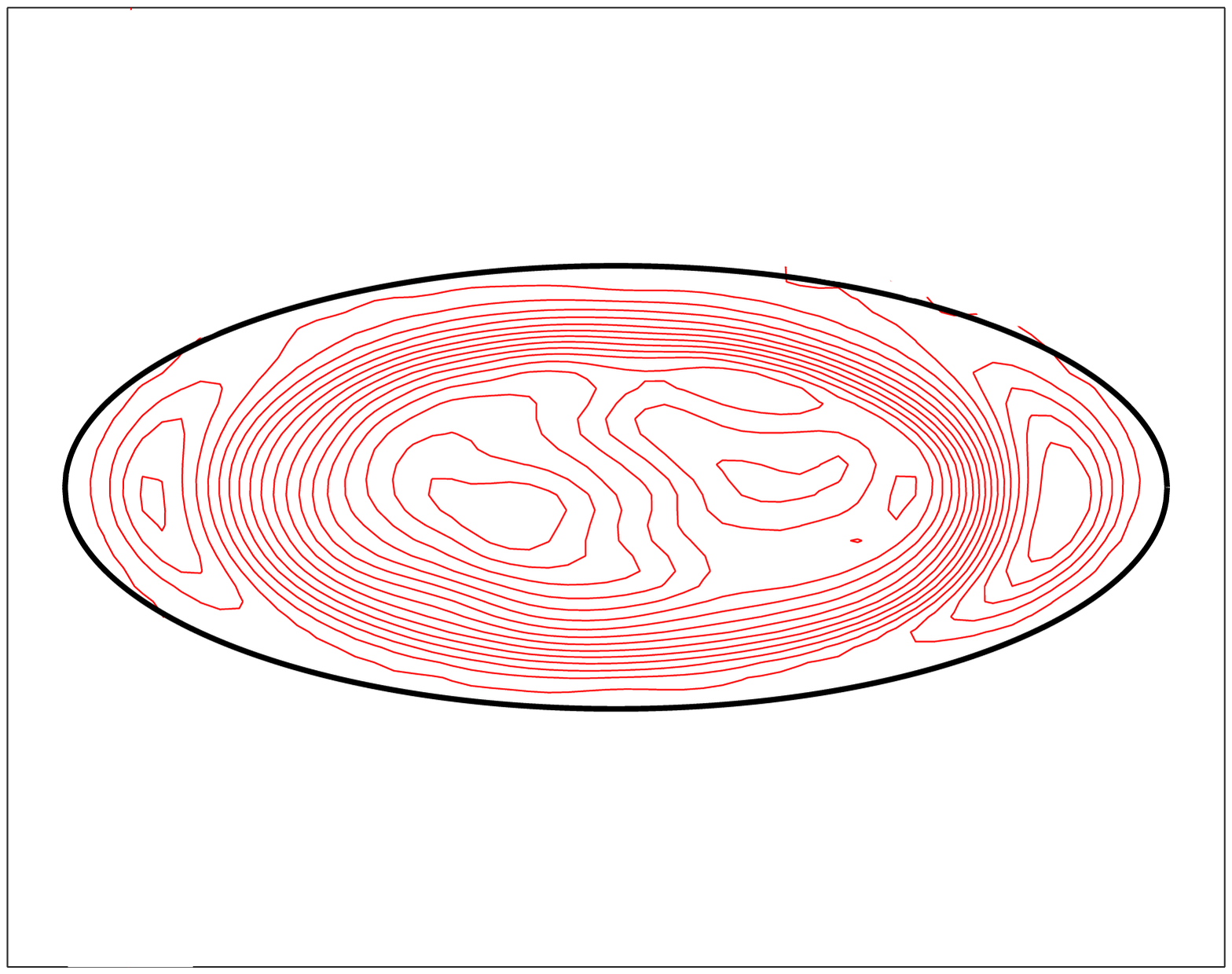}~\hspace{-0.6cm}
\caption{Flux-lines in a cross-section through the differently shaped geometries. From left to right the cross-sections are a circle with radius $a=1$, an ellipse with major semi-axis $a=1.2$ and one with $a=1.4$, for $S \approx 2\cdot 10^4$. Visualized is the induced field, i.e., the magnetic field without the imposed contribution. }
  \label{Fig:flux}
\end{figure}

\begin{figure*}
\includegraphics[width=.8\textwidth]{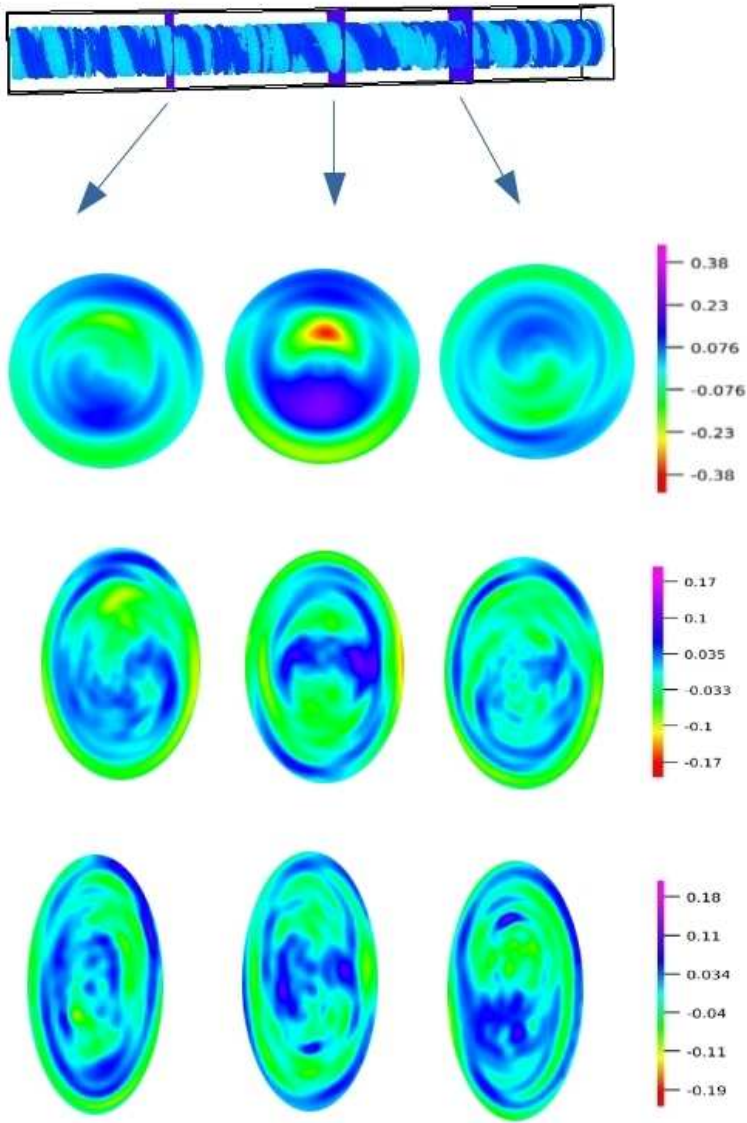}
\caption{Poloidal cross-sections of the cylinder, illustrating the axial magnetic fluctuations at $t=9.8.10^3 \tau_A$ for $S \approx 2\cdot 10^4$. The geometries are a circle with radius $a=1$, an ellipse with major semi-axis $a=1.2$ and one with $a=1.4$.
\label{Fig:PolModes}}
\end{figure*}

Classically in order to assess the influence of the pressure on the plasma dynamics, the quantity $\beta$ is evaluated, which measures the importance of the plasma pressure compared to the magnetic pressure 
\begin{equation}\label{eq:meanbeta}
\beta=2\frac{\langle P\rangle}{\langle B^2\rangle},
\end{equation}
where the brackets indicate a volume average over the plasma volume of interest. Other definitions are also used \citep{wesson2011tokamaks} based on the value of $\bm B$ at the wall of the plasma for instance.
Clearly ${\beta}$ is an important parameter since the plasma pressure is a key-parameter in the analysis of the confinement quality. However, if we are interested in the turbulence properties and the velocity field in general, it is the pressure-gradients which are important. It is known from turbulence-reasearch that estimating simply the order of magnitude of pressure gradients by $\mathcal O(\nabla P) \sim P/L$, with $L$ a macroscopic lengthscale can seriously underestimate their magnitude since the pressure gradients are in general dominated by small-scale contributions $l\ll L$. Therefore we recently introduced the alternative, gradient-based $\beta$s \citep{chahine2018role},
\begin{equation}
 \beta_\nabla\equiv2\frac{\langle\| \nabla P\|\rangle}{\langle\|\nabla  B^2 \|\rangle} \textrm{~~and~~} 
{\beta'_\nabla}\equiv\frac{\langle \| \nabla P \|\rangle}{\langle \| \bm J\times \bm B\|\rangle}.
 \end{equation}
It is only when ${\beta'_\nabla}$ is small {compared to unity} that the influence of the pressure term might be negligible in the dynamics, {compared to the influence of the other terms in the velocity equation}. This is not necessarily the case when $\beta$ is small.

In Fig.~\ref{fig:beta} we show the behavior of the different versions of $\beta$. The size of the temporal fluctuations of the different quantities around the time-averaged value are indicated by error-bars.
In agreement with the results in \citet{chahine2018role} the value of $ \beta$ is decreasing for increasing values of the Lundquist number. We see that for the highest values of $S$, $\beta$ has dropped most importantly in the circular geometry. This is not inconsistent with Fig.~\ref{fig:visu_beta}, which shows that the normalized pressure is weaker in this geometry than in the other two.
Indeed, focusing only on this observation one might be tempted to extrapolate and assume that the pressure plays no dominant role in the dynamics at high values of $S$. However, the gradient based $ \beta'_\nabla$ remains approximately constant when the Lundquist number is increased. This shows that the influence of the pressure-gradients does not become less important for larger $S$. 

These observations are quite robust, qualitatively independent on the shape of the cylinder. What changes however, is the $S$-dependence of ${\beta}_\nabla$. 
This indicates that $\langle\|\nabla  B^2 \|\rangle/ \langle \| \bm J\times \bm B\|\rangle$ is influenced by the ellipticity of the cylinder cross section. {From the vector identity $\bm J\times \bm B=-\nabla  B^2/2+\bm B\cdot \nabla \bm B $  this shows that the different behaviour is associated with the curvature term $\bm B\cdot \nabla \bm B$ which seems to be affected by the shaping.  } Shaping does therefore change the magnetic structure of the flow. We will further investigate this below, considering spectral considerations. Before that we will visualize the magnetic structure of the simulated plasmas.

\subsection{Characterization of the velocity and magnetic field}\label{sec:ResMod}

In Fig.~\ref{Fig:flux} we show magnetic flux lines in cross-sections through the three cylindrical geometries. The lines correspond to iso-values of the axial component of the magnetic vector potential. This potential is computed from the induced field, i.e., the imposed field is subtracted from the total magnetic field.

The most dramatic difference is observed for the $a=1.4$ geometry. Indeed, for the circular cross-section the iso-lines are mostly concentric. Changing the shape, for $a=1.2$ a separatrix appears at the edges of the ellipse and for $a=1.4$ the central flux surfaces are destroyed and magnetic islands appear.
 Clearly, the shape of the magnetic field is dramatically altered by the shaping in this latter geometry, compared to the circular cross-section.

In Fig. \ref{Fig:PolModes} axial magnetic fluctuations are shown in various poloidal cross-sections. The fluctuations of the axial magnetic field are obtained by showing the magnetic field without the axially invariant ($k_z=0$) magnetic contribution.  Only in a few sections a clear poloidal mode structure is observed. From these observations it seems relatively clear that no single-helicity state is present in our simulations. A large number of modes seems to be evolving simultaneously and instantaneous local visualizations are not the best tool to illustrate this.



\begin{figure*}
\subfigure[]{\includegraphics[width=.75\textwidth]{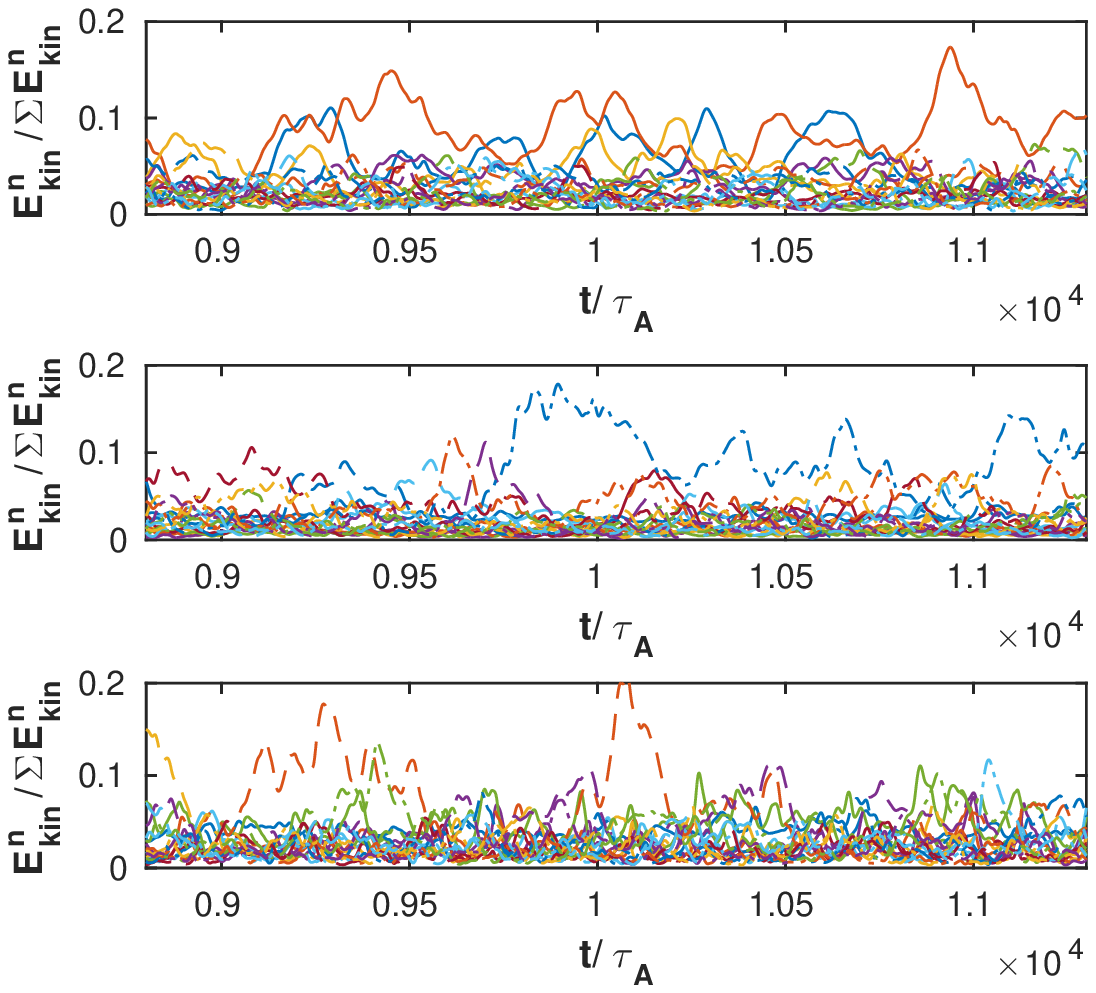}}~
\includegraphics[width=.17\textwidth]{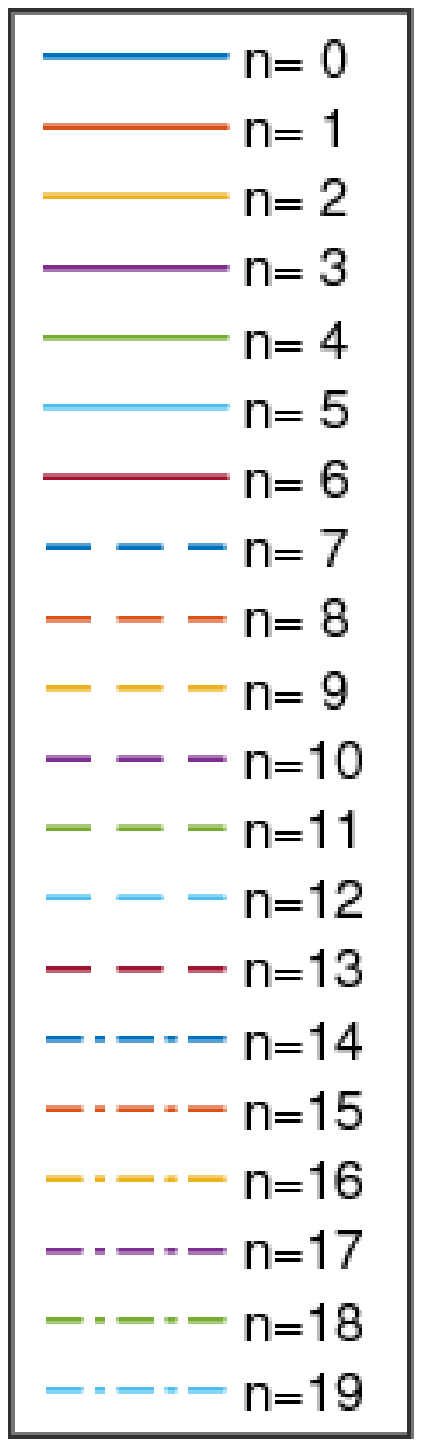}
\subfigure[]{\includegraphics[width=.75\textwidth]{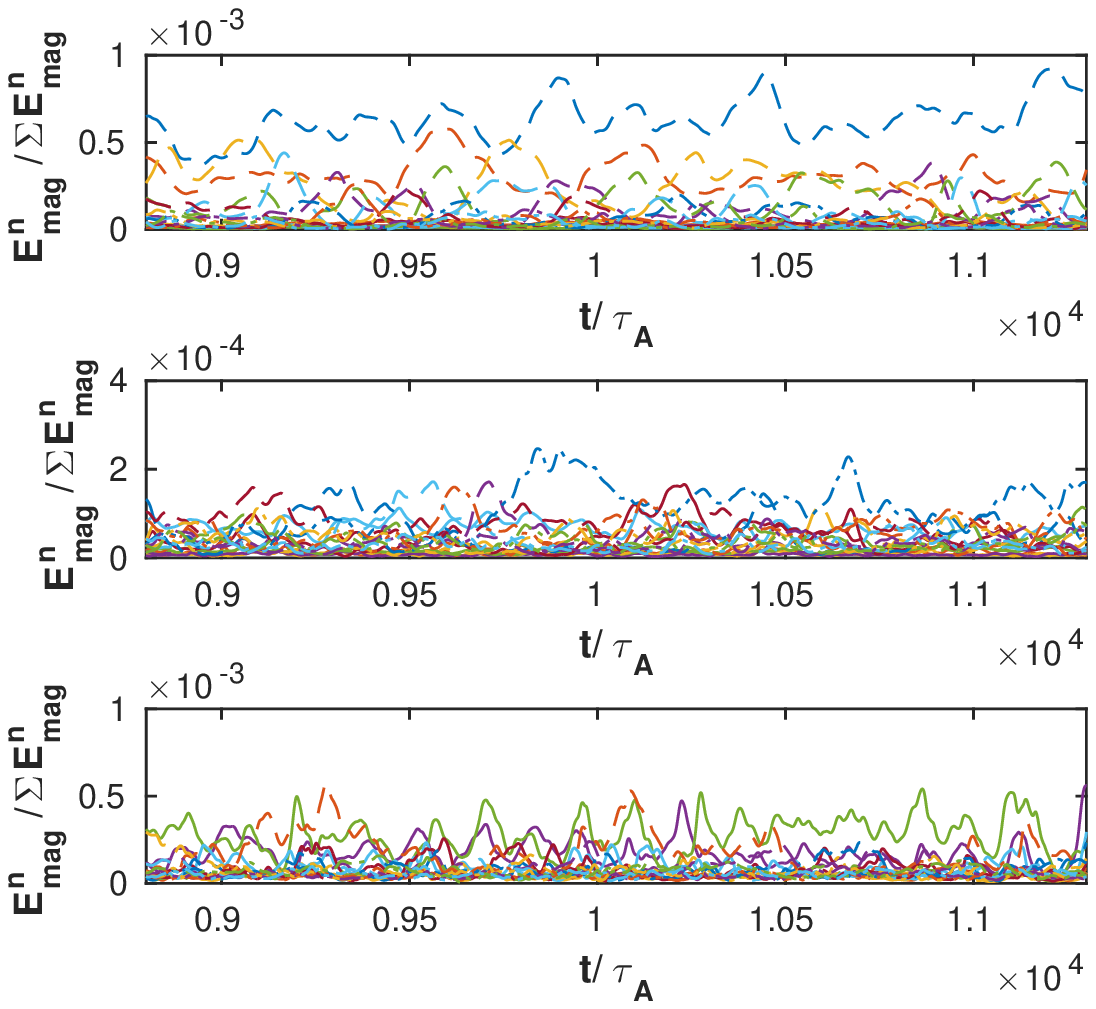}}
\caption{Axial spectra of kinetic (a) and magnetic (b) energy, normalized respectively by the total kinetic and magnetic energy. Three shapes are considered, respectively from top
 to bottom, a circle with radius $a=1$, an ellipse with major semi-axis $a=1.2$ and one with $a=1.4$, for $S \approx 2\cdot 10^4$.}
  \label{spectra_all}
\end{figure*}

To analyze the presence of different modes we visualize now the effect of shaping on these helical modes by evaluating the axially Fourier-transformed  magnetic field. In Fig.~\ref{spectra_all} we observe that all considered cases show strong fluctuations, illustrating that we are in a highly nonlinear regime, far away from quasi-static equilibria.
 
Fig.~\ref{spectra_all} shows  the predominance of a magnetic mode with toroidal mode number $n=7$ in the circular case, which is consistent with what has been observed in
the RFX-mod device \citep{lorenzini_2009,martin2009overview}. In the elliptical case ($a=1.2$) a tendency for mode $n=14$ to dominate is observed, while the magnetic modes $n=3$ and $n=4$ contain most of the magnetic energy for $a=1.4$.  This last case seems to be closer to a multiple-helicity state, where it is not a single mode which contains most of the energy. Similar spectral differences are observed in the kinetic spectra, where $n=0$ and $n=1$ are the dominating kinetic modes
in the circular case, $n=14$ in the $a=1.2$ elliptical case, and $n=8$ in the $a=1.4$ elliptical case.
A different representation of the modal dynamics is shown in the appendix where the modal spectra are plotted using double-logarithmic scale for the different cases.

 We have not carried out a detailed poloidal mode composition, which is in particular less convenient in the elliptical geometries, and we have therefore no further interpretation of the possible underlying instabilities which could be external tearing modes, or other MHD instabilities. We think that a linear instability analysis of the present system could allow further insights in the origin of the magnetic structures. However, since the modal spectra indicate the large range of active modes, the outcome of such an analysis is of limited use in the here considered fully turbulent state.

It is clear from these observations that shaping significantly influences the topology of the velocity and magnetic fields, both qualitatively and quantitatively. All three geometries do however show the presence of a large range of active scales.

\subsection{Poloidal angular momentum}

The presented results show a clear influence of the shape of the cross-section on the dynamics. It would be enlightening if we could understand these differences in the light of a clear, large scale dynamical feature which changes through the shaping. We have identified one such feature which behaves quite differently in the differently shaped geometries. This quantity is the poloidal angular momentum, a quantity which received considerable attention in two-dimensional flows.

In fact, an interesting difference between two-dimensional flows in circular or elliptical domains is the change in dynamics associated with a form of  symmetry breaking of the large scale flow patterns. Indeed in 2D turbulence, changing the flow-geometry from circular to elliptical, leads to the generation of angular momentum \citep{keetels_2008}. This effect was shown to persist in 2D MHD turbulence \citep{bos2008rapid,bos_2010} and its investigation is considered interesting in the context of confinement studies, since large scale poloidal motion could enhance radial transport barriers and stabilize MHD instabilities \citep{shan1994magnetohydrodynamic}. 

\begin{figure}
 \centering
  \includegraphics[width=.6\textwidth]{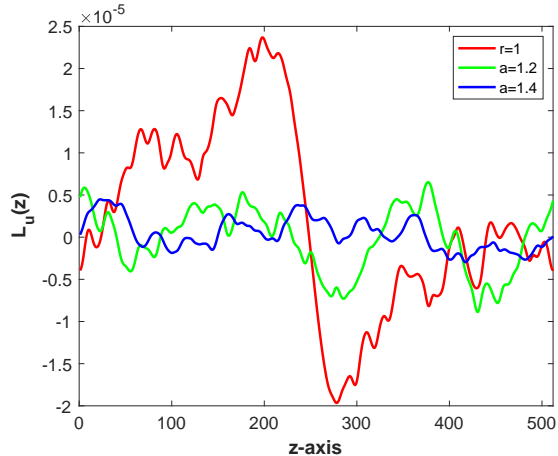}
 \caption{The $z$-dependence of the instantaneous poloidal angular momentum for the three geometries at $t=11\times10^3\tau_A$.}
 \label{ang_mom}
\end{figure}

An analysis of a similar possibility in the present geometry is shown in Fig.~\ref{ang_mom} where for a given time-instant the angular 
momentum associated with the poloidal flow is computed for each cross-section. Its definition is
\begin{equation}
L_u(z)=\int_A  \bm e_z\cdot(\bm r \times \bm u) \mathrm{d}A,
\end{equation}
where $\bm r$ is the radial vector, pointing from the center-line of the geometry outwards and $A$ is the surface of the poloidal cross-section. In a periodic cylinder with circular cross-section, global angular momentum can only be created by viscous effects. However, locally, at a given axial position the angular momentum can be strong. This is exactly what is observed for the case the circular geometry, where locally large values of the poloidal angular momentum exist. Through the associated rotating motion, confinement could be improved, since rotation suppresses instabilities \citep{shan1994magnetohydrodynamic}. Note that this is one particular time-instant, but qualitatively the same effect is observed at different time-instants, with local fluctuations of $L_u$, which are largest in amplitude in the circular geometry. {Possibly, this structure is associated with the $n=1$ velocity mode observed in the temporal spectrum of the kinetic energy in Fig.~\ref{spectra_all}(a).}

This observation is somewhat the opposite of what is observed in investigations of spontaneous spin-up in 2D systems. There it is seen that the generation of angular moment is absent in circular 2D domains and becomes more important in elliptical domains \citep{keetels_2008}. 

The important amount of circular movement characterized here by the poloidal angular momentum is perhaps simply due to the fact that a circular shape is more compatible with circular movement, compared to non-circular shaped cross-sections. However, such a poloidal flow could be beneficial for confinement, which shows here that with respect to that effect, circular domains might be favorable for good confinement. This is also what was concluded, for different reasons, in \citet{guo2013}. This observation is clearly of speculative nature and further research is required to investigate the link between poloidal angular momentum, shaping and the turbulent dynamics of the RFP.

\section{Conclusion}\label{sec:conc}

Direct numerical simulations of viscoresistive MHD show that in periodic cylindrical geometry in the RFP regime the shape of the 
cross-section significantly changes the nonlinear dynamics. Moreover, different helical states can be observed and different {toroidal} modes are excited in different geometries. Modifying the elliptical elongation leads to different modal behaviors. 

{We quantified first the influence of shaping on the $F-\Theta$ characteristics and showed that the influence of shaping on these global quantities is less important than a change in the Lundquist number. For other quantities, which do not concern the global properties, the picture is different. Indeed, shaping modified in our simulations significantly the magnetic flux-lines in a cross-section. It was observed that, for circular and moderately elliptical cross-section, the contours of the magnetic flux surfaces were roughly concentric. However, when the domain becomes elliptical enough ($a=1.4$ in our simulations), the contours started to show two magnetic islands. 

Perhaps most flagrant in our simulations is the generation of local poloidal angular momentum in the circular geometry, absent in the more elliptical shapes. The presence of such local, but large-scale poloidal structures could possibly stabilize the plasma dynamics. It seems, with respect to this feature, that circular cross-sections are better candidates to improve confinement. However, the addition of toroidal curvature to the system might alter this feature. As a matter of fact, in toroidal geometry a quasi-single helicity state was observed to be more persistent than in a cylindrical geometry \cite{morales_ppcf}. How poloidal flow-structures containing angular momentum and helical modes are related, and how they interact, deserves further attention.}


The most important outcome of this work is therefore not the determination of a certain value of the elongation, most efficient to obtain an optimal confinement, but the mere fact that elongation radically changes the dynamics of RFPs. We would therefore encourage experimentalists to consider the poloidal shape of the confining magnetic field as an important control parameter for RFP design and operation, since different modes are triggered when the shape of the cross-section is modified. If an experiment allows for a simple modification of the plasma shape, it might give more freedom to obtain a competitive fusion plasma.

\section*{Data availability}

The data that support the findings of this study are available from the corresponding author
upon reasonable request.

\section*{Acknowledgments}
The authors acknowledge the help and comments of Jorge A. Morales and discussion with D. Bonfiglio, D.F. Escande and M. Veranda. We further acknowledge support by the French Federation for Magnetic Fusion Studies (FR-FCM) and of the Eurofusion consortium from the Euratom research and training programme 2014–2018 and 2019–2020 under grant agreement No633053. The views and opinions expressed herein do not necessarily reflect those of the European Commission.
We acknowledge IDRIS (Project No. 22206), PMCS2I and P2CHPD for the use of their facilities.  

\section*{Appendix}

\subsection*{Temporal evolution of modal spectra}

\begin{figure*}
\subfigure[]{\includegraphics[width=0.4\columnwidth]{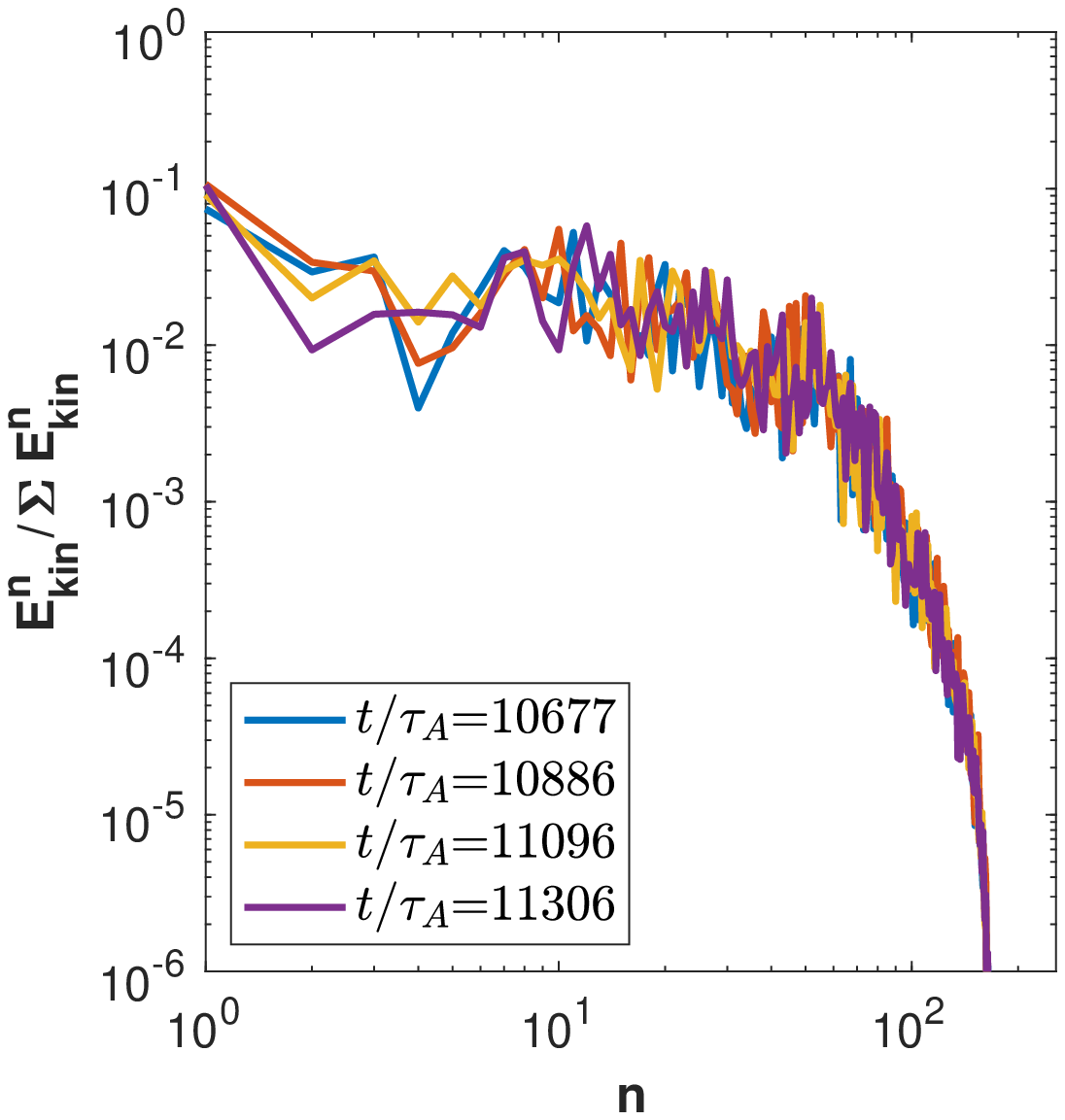}}
\subfigure[]{\includegraphics[width=0.4\columnwidth]{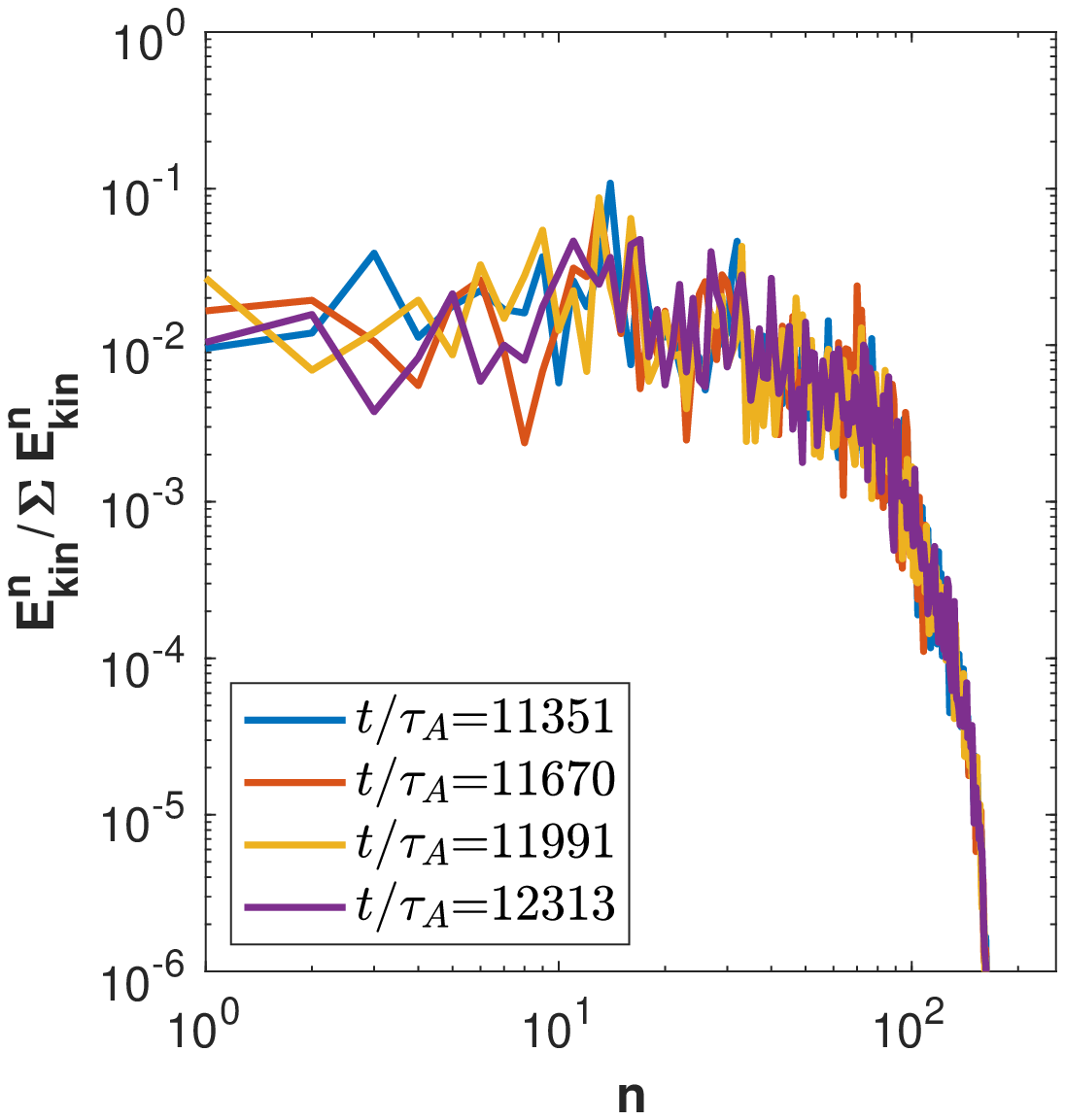}}
\subfigure[]{\includegraphics[width=0.4\columnwidth]{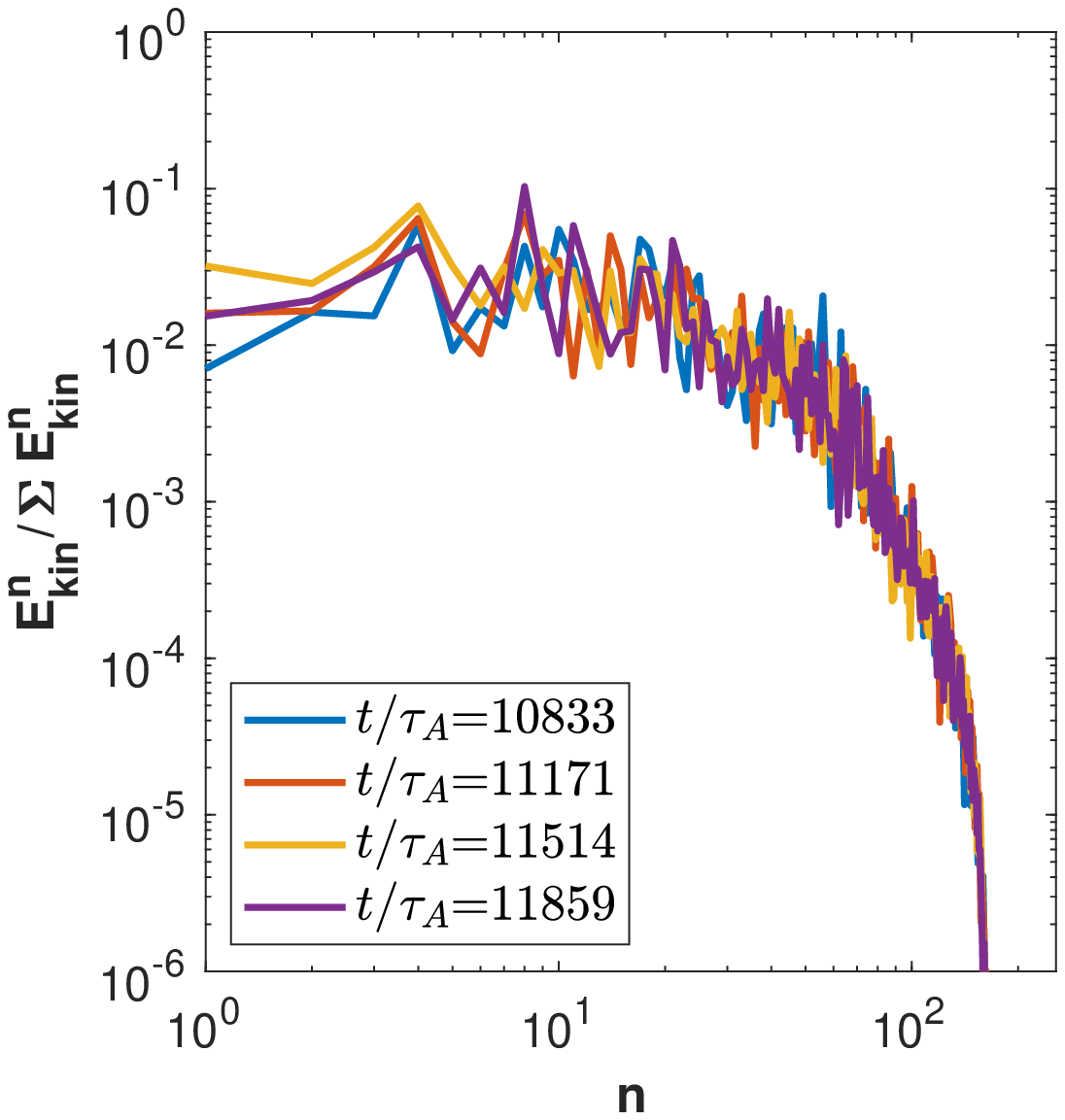}}
\subfigure[]{\includegraphics[width=0.4\columnwidth]{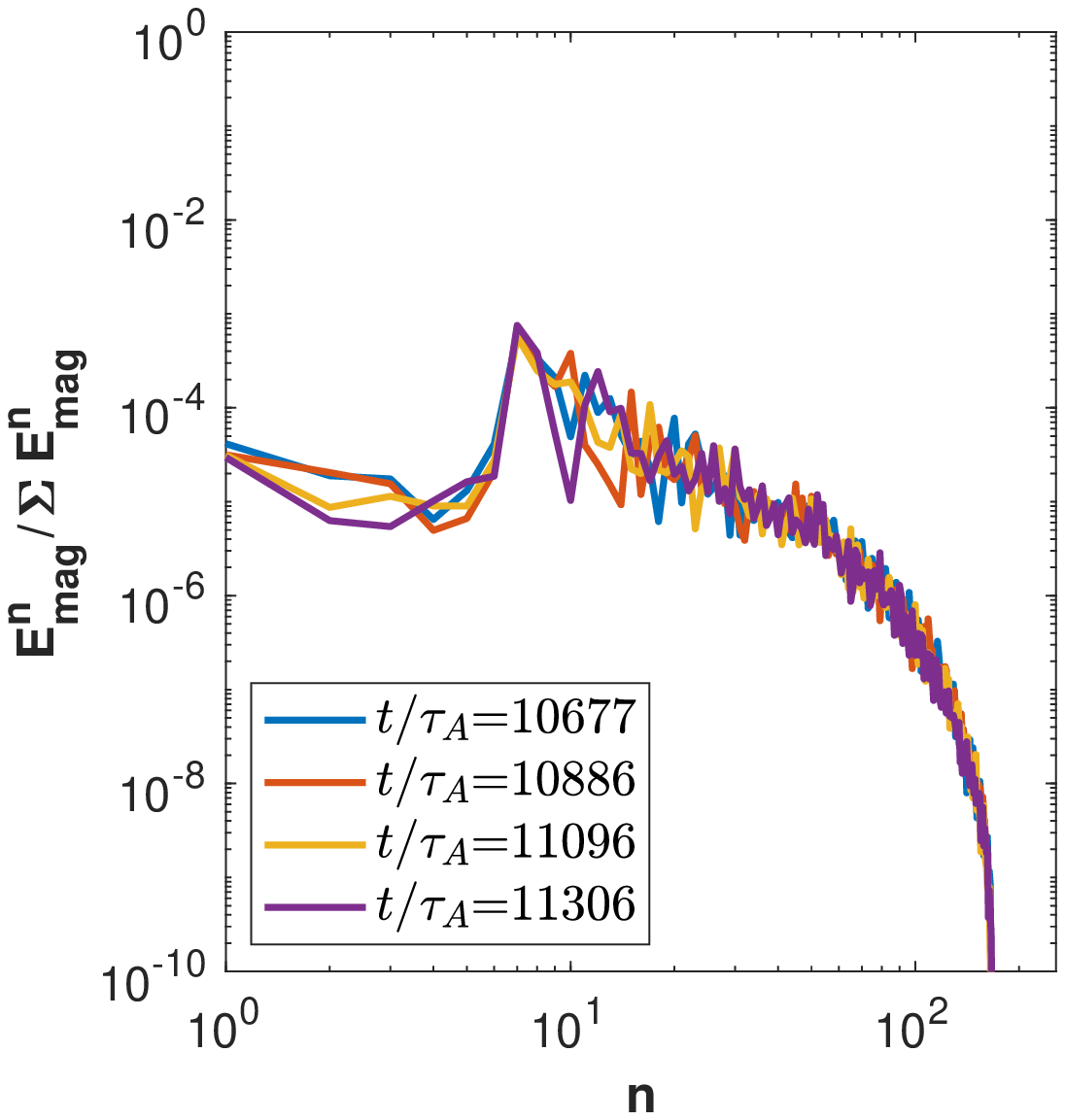}}
\subfigure[]{\includegraphics[width=0.4\columnwidth]{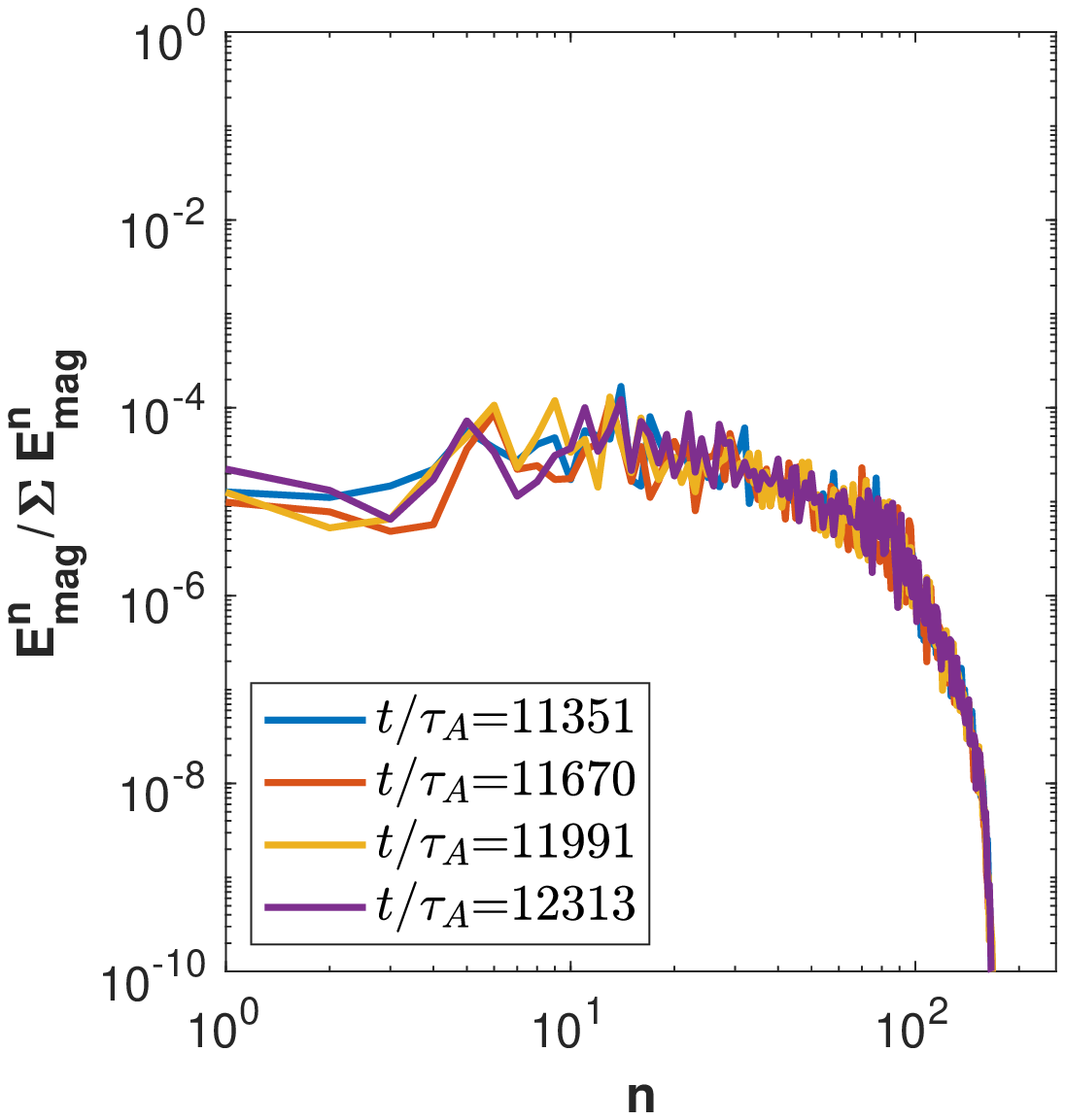}}
~~~~~~~~~~~~~~~~~~~~~~~
\subfigure[]{\includegraphics[width=0.4\columnwidth]{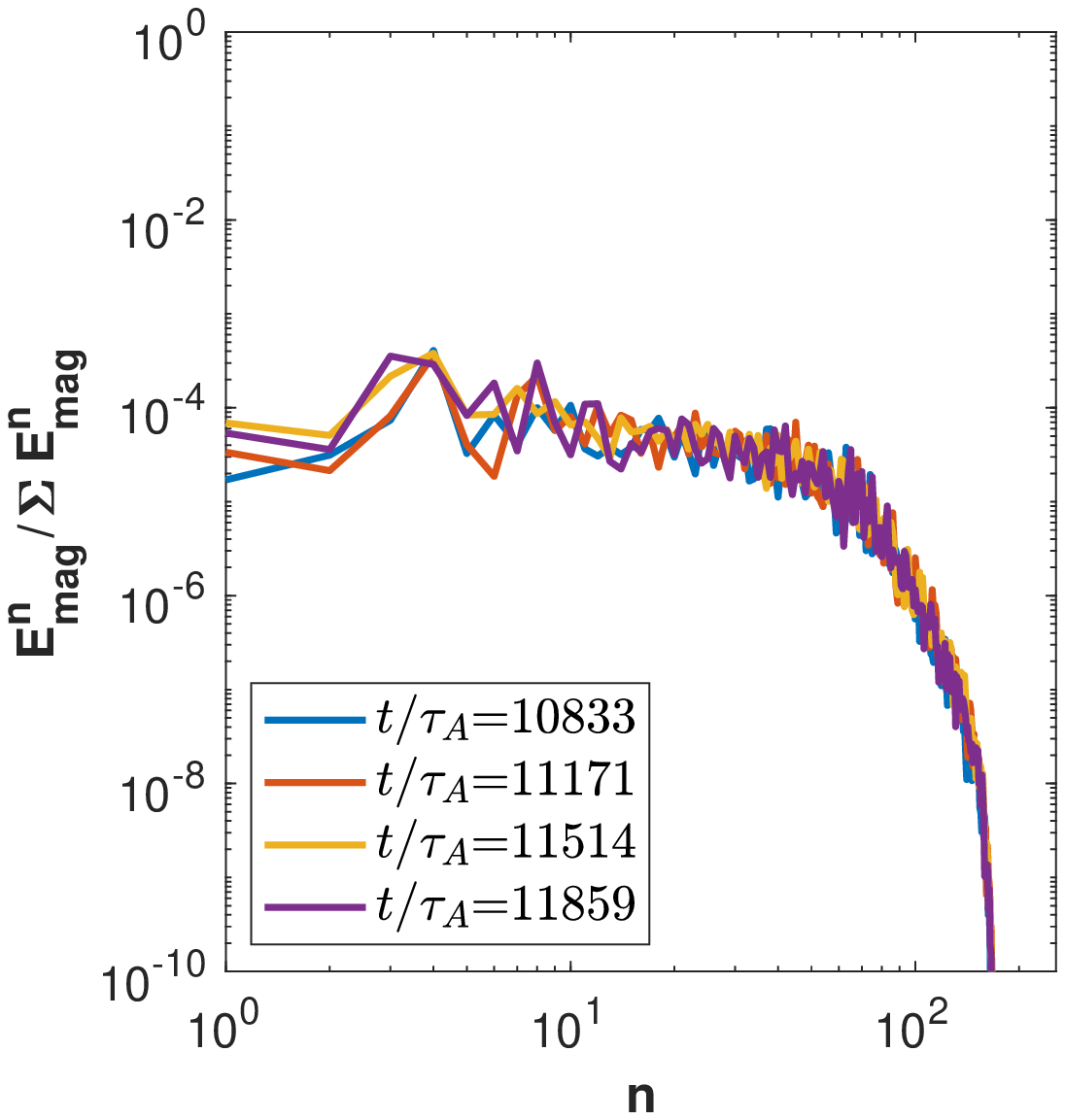}}
\caption{Axial mode spectra of kinetic energy fluctuations (top row) and magnetic fluctuations (bottom row) in cylinders with ellipticity (a,d) $a=1$, (b,e) $a=1.2$, (c,f) $a=1.4$. Spectra are shown at four different time-instants during the statistically steady state. The Lundquist number is $S\approx 2\cdot 10^4$. 
 \label{fig:specsub}}
\end{figure*}

To complete the modal analysis carried out in Sec.~\ref{sec:ResMod}, we show in Fig.~\ref{fig:specsub} the modal spectra in double-logarithmic representation. For each spectrum four time-instants are shown during the statistically steady state. The most striking observation is the large number of active modes. In this representation we cannot clearly identify a dominant mode in the system. Quasi-single helicity states might be attainable, but this is certainly not observed in the present results.

\bibliographystyle{jpp}

\bibliography{references}

\end{document}